\documentclass[10pt]{wlscirep}
\usepackage[utf8]{inputenc}
\usepackage[T1]{fontenc}
\usepackage{lineno}
\usepackage{graphicx}
\usepackage{upgreek}

\title{Programmable Space-Frequency Linear Transformations in Photonic Interlacing Architectures}

\author[1,2]{Jonathan Friedman}
\author[1]{Kevin Zelaya}
\author[1,2]{Mostafa Honari-Latifpour}
\author[1,2,*]{Mohammad-Ali Miri}
\affil[1]{Department of Physics, Queens College of the City University of New York, Queens, New York 11367, USA}
\affil[2]{The Graduate Center of the City University of New York, New York, New York 10016, USA}
\affil[*]{Corresponding author (mmiri@qc.cuny.edu)}

\begin{abstract} 
Programmable photonic circuits are versatile platforms that route light through multiple interference paths using reconfigurable optoelectronic elements to perform complex discrete linear operations. These circuits offer the potential for high-speed and low-power photonic information processing in various applications. The mainstream research on programmable photonics has focused on implementing linear operations on discrete signals encoded in the modal amplitudes of an array of spatially separated single-mode waveguides. However, many photonic device applications require simultaneous transformations in the space-frequency domain, where information is encoded in both the spatial modes of waveguides and their spectral content. Here, we experimentally demonstrate linear space-frequency transformations using a $4 \times 4$-port programmable silicon photonic circuit with an alternating architecture. This design leverages the limited dispersion of coupled waveguide arrays to enable linear operations with reconfigurable frequency-dependent matrix elements. We utilize this device to perform wavelength demultiplexing and filtering. This architecture platform can pave the way for versatile devices with applications ranging from wavelength routing to programmable dispersion control.

\end{abstract}

\begin{document}
\flushbottom
\maketitle
\thispagestyle{empty}

\section{Introduction}

The exploration of optical phenomena for computational tasks has been a research-intensive and evolving field, tracing back to the free-space solution based on lenses for performing Fourier transforms~\cite{Goodman05} and progressing to the realization of unitary operations with beam splitters and mirrors~\cite{Reck94}. Although the scalability of early implementations was limited due to bulky and vibration-sensitive elements, these efforts paved the way for more compact and efficient solutions. Rapid advances in nanotechnology over the past few decades~\cite{yang2025nanofabrication} have enabled the fabrication of optical structures that are, in turn, capable of performing complex mathematical operations by exploiting the fundamental properties of light propagation~\cite{joannopoulos1997photonic,mcmahon2023physics,yan2009nanowire}. In this regard, on-chip programmable photonic linear transforms are of particular interest, as they create new opportunities in quantum information and quantum transport simulations ~\cite{madsen2022quantum,Harris17,slussarenko2019photonic}, optical signal processing \cite{notaros2017programmable}, neuromorphic computing~\cite{Xu23neuromorphic}, optical neural networks~\cite{shen2017deep,zhu2022space,li2024high,yang2024complex,zelaya2025integrated}, and enable rapid prototyping of linear multiport photonic devices~\cite{Tang:22,xu2022parallel}. 


Numerous techniques have been investigated in the literature to realize unitary linear transformations within compact on-chip photonic circuits. These include photonic architectures employing networks of Mach-Zehnder interferometers (MZIs), such as the rectangular Clements mesh~\cite{Clements16}, the diamond mesh~\cite{shokraneh2020}, and other configurations~\cite{Mojaver2023}. Additionally, topological photonic lattices featuring hexagonal MZI arrays provide another alternative approach~\cite{Wang2019,on2024programmable}. Recent research has explored interleaved architectures that combine passive transfer matrices with tunable active phase layers~\cite{Tanomura20, Saygin2020, Tanomura22a, Pastor21, markowitz2023universal, zelaya2024goldilocks}, where the active layer is implemented through tunable phase elements. A particular approach utilizes multimode slab waveguides to perform discrete Fourier transforms (DFT), achieving universal transformations when a sufficient number of layers are incorporated~\cite{Pastor21}. Metasurface-based structures have also emerged as a promising avenue for optical computing, enabling computational functionalities for large-scale~\cite{tzarouchis2025programmable} and component-level devices through inverse-design approaches~\cite{piggott2017fabrication,molesky2018inverse}. Alternative devices exploit the polarization degrees of freedom in multilayer structures~\cite{Tanomura23}, and multiple frequencies have been explored by embedding resonators in the structure~\cite{buddhiraju2021arbitrary}. 

Recent studies have found rigorous numerical evidence that interleaved phase arrays and discrete fractional Fourier transform (DFrFT) layers render universal $N$-dimensional unitary transforms with only $N+1$ layers~\cite{markowitz2023universal}. A further generalization using Haar random unitaries as the interlacing layer has yielded the desired universality as long as the passive layer satisfies a density criterion~\cite{zelaya2024goldilocks}. In our earlier work, we proved that the interlacing architectures are robust to fabrication defects and exhibit auto-calibrating properties~\cite{Markowitz23Auto}. Extensions to non-unitary optical computing have risen as a natural extension, such as the singular value decomposition in networks of MZIs~\cite{Miller2012,moralis2024perfect}, interlaced structure combined with amplitude modulation layers~\cite{markowitz2024learning}, and embedding lower-dimensional matrices in higher-dimensional optical linear transformers~\cite{markowitz2025embedding}. 

This work introduces a programmable photonic integrated circuit (PIC) with enhanced capabilities to operate in the wavelength domain. The PIC design consists of $4\times 4$ passive layers created using dispersive coupled waveguide arrays, along with active layers featuring phase shifters. Although individual waveguide arrays exhibit a low dispersive response, their combined effect yields a richer wavelength response. The phase shifters allow for parameterization of the transmission matrix, whose components are trainable and wavelength-dependent. 
The device is fabricated as a four-port system on a silicon-on-insulator platform, integrating thermo-optic microheaters for active phase control and optimized to operate within the C-band. The fabricated chip is both optically and electrically packaged, interfacing with optical and electrical control units that are operated through a computer-assisted framework for real-time configuration. 

Unlike passive solutions for unitary control, such as inverse-designed units, the proposed device enables on-the-fly programming, allowing for corrections due to fabrication defects. The device supports space-frequency operations, which are enhanced through an in-situ training approach that employs optimization algorithms, such as gradient descent, hill climbing, and simulated annealing to evaluate the convergence of various target functions for both single and multiple wavelength operations.

\section{Results}

\subsection{Theory and design}
The proposed PIC is mathematically represented by a parameterized unitary operator $\mathcal{U}({\Phi}) \in U(N)$, where ${\Phi}$ denotes a set of tunable parameters used to induce optical interference and steer light to the desired output distribution. Due to the nature of the mixing layer, this device ideally operates as a lossless device. The unitary nature of the architecture is particularly suitable for implementation in optical systems governed by coupled-mode theory. In such systems, guided modes propagate through evanescently coupled dielectric waveguides, with their evolution described by unitary wave dynamics~\cite{yariv2007photonics,Huang94}. In this context, for single-mode waveguides, the electric field of an excitation propagating through the waveguide array along the $z$-direction can be expressed as $E(x,y,z)=\sum_{n=1}^{N}\mathcal{E}(x,y-y_{n})a_{p}(z)$, where $\mathcal{E}(x,y-y_{n})$ is the unit-power normalized electric field of the guided-mode, and $a_{p}(z)\in\mathbb{C}$ denotes the corresponding complex-valued mode amplitude of the $p$-th waveguide. The mode amplitudes govern the dynamics of the electric field in the structure, defined through the equation $-id\mathbf{a}(z)/dz=H\mathbf{a}$ with $\mathbf{a}(z):=(a_{1}(z),\ldots, a_{N}(z))^{T}$ and $H\in\mathbb{C}^{N\times N}$ a tridiagonal tight-binding Hamiltonian that dictates the coupling among waveguides~\cite{yariv2007photonics}. For the structures under consideration, the Hamiltonian can be considered as Hermitian and independent of the propagation distance $z$, which allows for a simple unitary evolution of the form $\mathbf{a}(z)=F(z)\mathbf{a}(z=0)$, with $F(z)\equiv F = e^{-izH}$. The latter highlights the matrix-vector operation performed by the waveguide array on an initial excitation $\mathbf{a}(z=0)$, which is the cornerstone principle exploited throughout the rest of the manuscript.

The proposed device comprises four input and output ports designed to perform space-frequency transformations on optical signals fed into the device inputs. The device is designed with an interlacing structure~\cite{Markowitz23Auto, markowitz2023universal}, which exploits the unitary nature of dispersive waveguide arrays $F$ (passive layers), yielding wavelength-dependent light mixing across the different waveguides. Indeed, dispersion on such structures is known to be limited, with a flat response. However, the combined action of all these couplers induces richer behavior, allowing for wavelength-dependent tunability. See the experimental setup section for more details. The structure is completed by interspersed phase shifters $e^{i\Phi^{(k)}}$ (active layers) that operate as the tunable elements in the structure, rendering a programmable solution. The proposed structure is summarized in the block diagram shown in Figure~\ref{fig:schematic}\textbf{a}. Thus, the overall wave evolution in the device is described by the unitary propagator
\begin{equation}
\mathcal{U}({\Phi}) = F e^{i\Phi^{(5)}} F e^{i\Phi^{(4)}} \ldots e^{i\Phi^{(2)}}F e^{i\Phi^{(1)}}F,
\label{eqU}
\end{equation}
where each $\Phi^{(k)} = \operatorname{diag}(\phi_{1}^{(k)}, \ldots, \phi_{N}^{(k)})$ is a diagonal matrix representing a tunable phase-shifters in the k-th layer (active layer), and $F \in U(N)$ is a fixed unitary matrix characterizing the waveguide array (passive layer). For the device discussed throughout this manuscript, we consider a four-port device with $M=5$ layers of active elements, where each layer comprises $N=4$ tunable elements, resulting in a total of $20$ active parameters in the architecture~\eqref{eqU}. The universality of such a device has been numerically assessed for different classes of waveguide arrays and other unitary couplers~\cite{zelaya2024goldilocks}. Indeed, experimentally achieving the theoretically predicted transfer function $F$ of the passive coupler is limited by fabrication constraints, and thus, defects are unavoidable. Nevertheless, the interlace design is robust against such defects, which can be further compensated by adjusting the phase shifters without compromising the required universality~\cite{Markowitz23Auto}. Adaptability is one of the main features that ensures the experimental viability of the proposed design.

\subsection{PIC layout design and fabrication}
\begin{figure}[t!]
    \centering
    \includegraphics[width=0.8\textwidth]{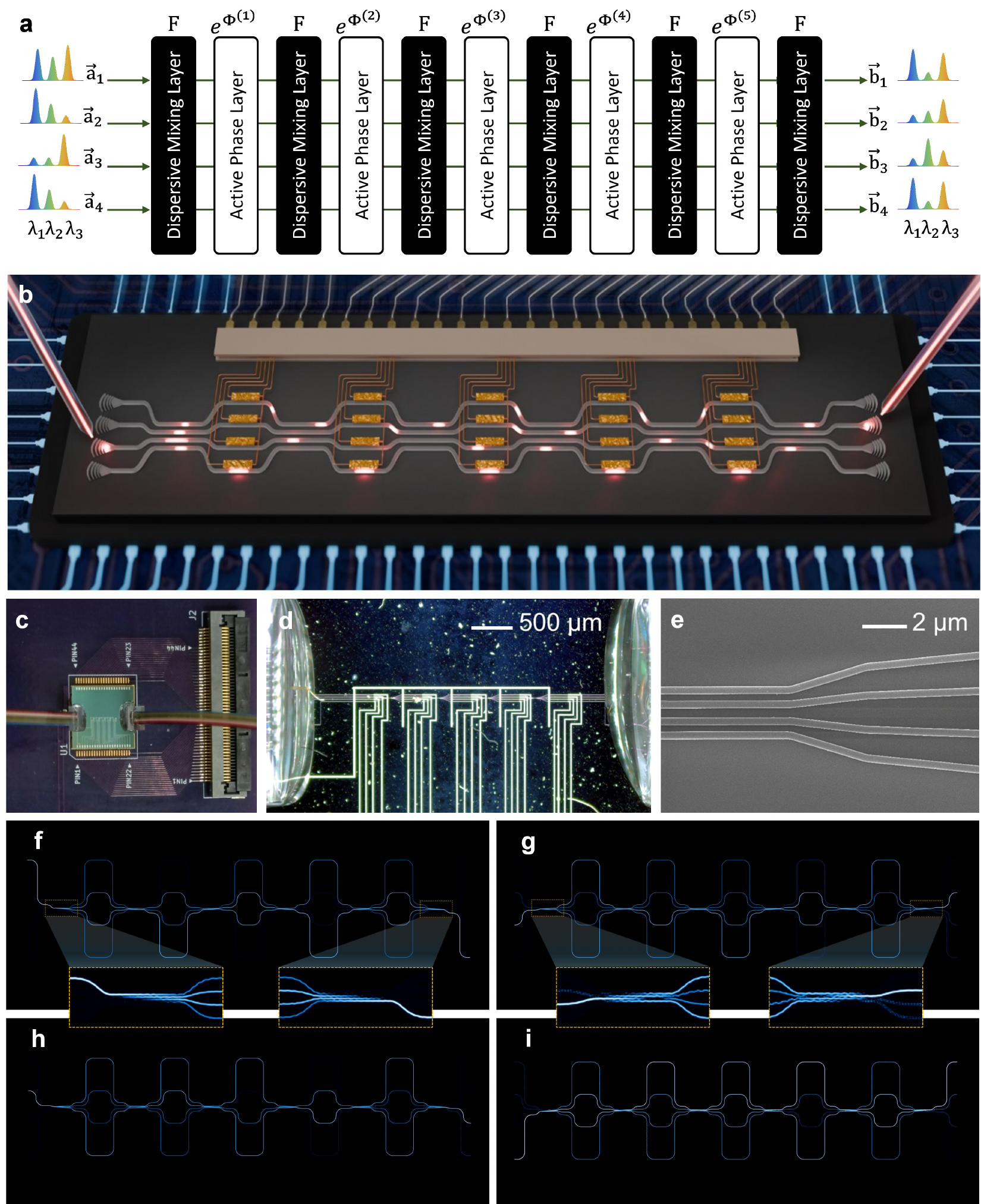}
    \caption{\textbf{Reconfigurable programmable photonic integrated circuits design}. \textbf{a} Block diagram of the proposed $4\times 4$ unitary and programmable device. This design features low-dispersive mixing layers that work together to create a richer dispersive response. \textbf{b} Schematic representation of the proposed PIC, illustrating the dispersive waveguide arrays (passive units) and the phase shifters (active units) utilized in the interlacing structure. \textbf{c} Image of the as-fabricated packaged chiplet on the PCB board assembled with the optical fiber array and electrical wiring. \textbf{d} Image of the fabricated PIC. Lines starting near the bottom are the electrical traces connecting the metal heaters. \textbf{e} SEM micrograph of a zoomed-in view of the four-port waveguide array section. \textbf{f}-\textbf{i} Electromagnetic wave simulations of the proposed PIC. The phase elements have been simulated by changing the effective index of the waveguide arms to introduce desired phase changes. The phases were tuned to achieve the cross configuration, which routes the excitations in the input ports $\{1,2,3,4\}$ to the output ports $\{4,3,2,1\}$, respectively.}
    \label{fig:schematic}
\end{figure}

The current device is fabricated according to the interlacing structure depicted in Figure~\ref{fig:schematic}\textbf{b}. The passive layers are formed by identical waveguide arrays, the geometries of which are chosen to support a unique guided mode through each waveguide. This is achieved by fabricating the waveguides on a silicon-on-silica platform, which comprises a silicon (Si) core material with cross-sectional dimensions of 500 nm x 220 nm, buried in a silica cladding. The high-contrast refractive index ensures that a single mode is supported in this geometry for an operational wavelength of 1550 nm, thus minimizing artifacts due to mode crosstalk. The waveguide array is designed, simulated, and optimized by determining the optimal separation between neighboring waveguides and the coupling length that yields maximum mixing. That is, the transfer matrix of the coupler fulfills the density requirement, a property that theoretically ensures universality on the proposed PIC~\cite{zelaya2024goldilocks}. The transfer matrix of the fabricated sample does not need to precisely match the predicted one, since deviations due to manufacturing defects can be compensated by recalibrating the tunable phase elements~\cite{Markowitz23Auto}. 

In turn, the active layer incorporates electrically tunable phase shifters that serve as training parameters in the in-situ training operations. Metal alloy microheaters implement a thermo-optical effect, allowing a phase shift of at least $\pi$. To prevent thermal or electrical crosstalk, thereby ensuring isolation between neighboring waveguide modes, the microheaters are strategically placed within the extended waveguide layer. The chip includes six waveguide arrays and five phase layers, rendering a total of 20 microheater elements, of which only 17 are functional due to broken wiring in three pads on the PCB board during handling in our lab. As shown in previous works~\cite{Markowitz23Auto}, this design is resilient to manufacturing defects; therefore, the broken metal heaters should not significantly impact device performance. See the experimental results section below for details.

Since our solution follows a simple design compatible with open-access foundries, the entire PIC layout was designed according to the previous prescriptions and submitted for fabrication to Applied Nanotools Inc. (ANT) using their silicon-on-insulator (SOI) technology Multi-Project Wafer (MPW) run. The fabricated chip was optically and electrically packaged by ANT, where the optical packaging included coupling and bonding of V-groove fiber arrays featuring an 8-degree polish angle and 250 $\mathrm{\mu m}$ pitch. In turn, electrical packaging was accomplished by wire bonding to the electrical pads on a PCB. The packaged PCB is shown in Figure~\ref{fig:schematic}\textbf{c}, a close-up of the PIC is shown in Figure~\ref{fig:schematic}\textbf{d}, and an SEM micrograph of the coupling region in the coupled waveguide arrays is depicted in Figure~\ref{fig:schematic}\textbf{e}.

Before manufacturing the device, we validate the layout design and its functionality through electromagnetic wave simulations using COMSOL Multiphysics. An individual waveguide array was initially simulated and characterized to ensure it produced sufficient mixing throughout all the waveguides. This waveguide array was then replicated and interconnected with the rest of the PIC using adiabatically bent waveguides. Additionally, circular bends with a radius of 15 $\upmu$m were employed to route light to the phase shifter section. The phase elements were simulated by adjusting the effective refractive index of the horizontal arms of the waveguide to induce the desired phase shift. The phases were optimized to produce an anti-diagonal identity matrix (a type of permutation matrix) so that light entering the excitation ports {1, 2, 3, 4} is redirected to the output ports {4, 3, 2, 1}, respectively. Simulation results for the cross configuration are presented in Figure~\ref{fig:schematic}\textbf{f}-\textbf{i}. These results illustrate how light from a single excitation propagates through the structure and ultimately emerges at the desired output. Although in some instances, such as the right inset in Figure~\ref{fig:schematic}\textbf{h}, a small amount of power is directed to output ports other than port 2, the normalized output transmittance at the desired port is greater than 0.98. The insets illustrate the mixing nature of the waveguide array, which distributes power evenly among all waveguides for single-port excitations. 

\subsection{Experimental setup and in-situ training}
\begin{figure}[t!]
    \centering
    \includegraphics[width=0.85\textwidth]{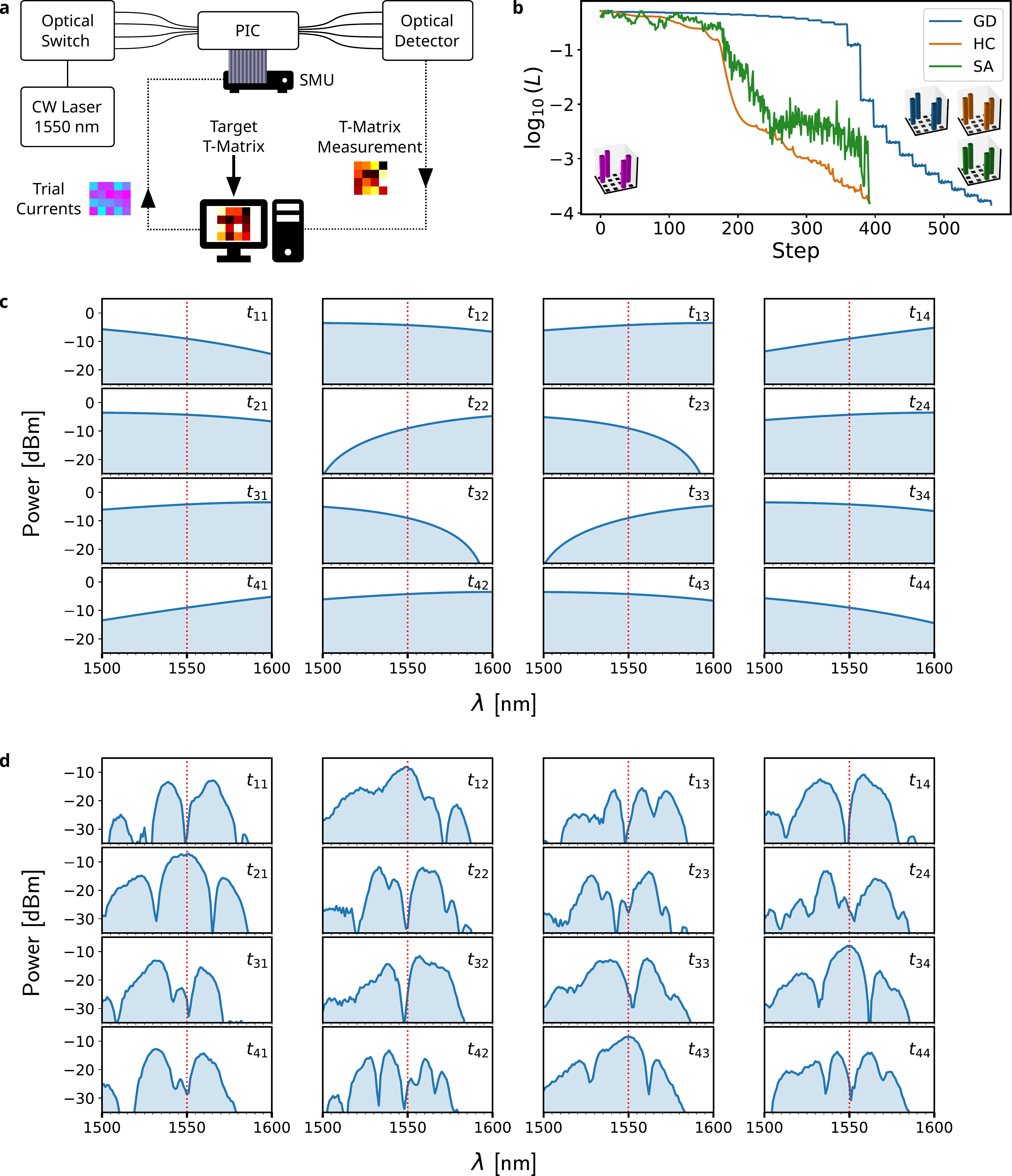}
    \caption{ \textbf{Experimental setup for in-situ training and wavelength-sweep measurements.} \textbf{a} The measurement system consists of a laser source, an optical switch, polarization paddles, and an electrical source-measure unit (SMU) connected to the PIC to drive both optical and electrical signals. Starting with an initial random set of current values, the system generates a transmission matrix from the chip and then minimizes the error relative to the target matrix. The computer program automatically continues iterating until the error drops below a predefined threshold. \textbf{b} Simulated wavelength response of the coupled waveguide array used as the dispersive mixing layer ($F$) in the device. \textbf{c} Optimization error versus step for the target matrix shown in the lower-left inset. The transmission matrices obtained after various optimization methods are shown in the right insets. \textbf{d} Wavelength-sweep of system optimized for identity matrix permutation 2143 at 1550 nm.}
    \label{fig:architecture}
\end{figure}

The block diagram in Figure~\ref{fig:architecture}\textbf{a} illustrates the experimental setup used to demonstrate the functionality of the programmable PIC through in-situ training. The fully packaged chip system is connected to both optical and electrical source-measure units and operates via a Python script. In this setup, the input fiber array is linked to a tunable laser through a 4-channel optical switch network and polarization controllers, while the output fiber array is directly connected to a 4-channel optical detector for recording the real-time power of the manipulated light. The electrical current to each phase shifter is supplied by an electrical source-measure unit (SMU) via an FPC connector on the PCB board.

Optimization methods can involve either in-situ or ex-situ training processes, depending on the complexity and specific needs of the application. For the current device, we developed an in-situ training approach for our proposed chip to calibrate the complex-valued target function in real-time. In the training network, the optimization process involves specifying an objective function to be minimized. We construct the figure of merit based on the number of input and output ports, as well as the number of frequency points. The frequency points are discretely sampled around the C-band. Therefore, the figure of merit is computed from the desired target matrix $T^{g}(\lambda)\in\mathbb{C}^{N_{1}\times N_{2}}$ and the measured matrix $\mathcal{T}^{m}(\Phi,\lambda)$ through the semi-positive definite function
\begin{equation}
\mathcal{L}(\Phi;\Lambda) := \frac{\Sigma_{\lambda \in \Lambda} \Vert \mathcal{T}^{m}(\Phi;\lambda) - T^{g}(\lambda) \Vert^2}{N_{1} N_{2} N_\lambda}, \quad \Lambda := \{{\lambda_{1}, \lambda_{2}, ..., \lambda_{N_\lambda}}\},
\label{L_multi}
\end{equation}
where $\Vert \cdot \Vert$ represents the Frobenius norm ($L^2$ norm), and $N_\lambda$ is the number of frequency channels. $N_{1}$ and $N_{2}$ are the number of rows and columns, respectively, of the target matrix. The measured matrix $\mathcal{T}^{m}(\Phi;\lambda)$ updates by manipulating the active parameter set $\Phi$, which is tuned by the currents applied to the microheaters in the structure. In some configurations, the optimization is carried out by exciting only a single port and sweeping different wavelengths, resulting in $N_{2}=1$ and $N_{\lambda}>1$. For operations at a single frequency, we set $N_{\lambda}=1$ and set $N_{1}=N_{2}=4$ to characterize the transfer matrix of the device.

When addressing arbitrary targets, we expect numerous local minima in the optimization landscape defined by the figure of merit, making it challenging to reach the global minimum. The selection of optimization algorithms depends on the nature of the problem. In this case, we have utilized selective optimization techniques—such as gradient descent, hill climbing, and simulated annealing—each of which is effective and well-suited for fine-tuning the system to deterministically identify global minima and achieve convergence toward the target function. 

In each iteration of the algorithm, the current values of all heater elements are automatically adjusted to drive the figure of merit $\mathcal{L}(\Phi)$ to a predefined threshold limit. In the class of programmable devices under investigation, the number of waveguide arrays and phase shifter layers is a crucial factor for conducting prompt convergence. Parameters for the various optimization algorithms include the initial applied currents, error-threshold limits, step sizes, and iteration boundaries. To further test the training capabilities of the fabricated PIC, we consider the representation of the whole permutation group in $\mathbb{C}^{4}$, which forms a subgroup of $U(4)$. The latter contains $4!$ matrices in total, with all entries zero except for a total of four nonzero components, where each nonzero component is uniquely positioned such that there is only one nonzero entry per row and column. The permutation matrices are used as the targets $T^{t}$ at a single wavelength of 1550 nm during the training process. 

Figure~\ref{fig:architecture}\textbf{b} depicts the convergence rate by using the gradient-descent, hill-climbing, and simulated annealing algorithms for the permutation matrix 2143 as the target at the 1550 nm wavelength. Indeed, permutation matrices are prime examples of unitary sparse matrices, which are compatible with device capabilities. The experimentally measured normalized transmission is shown in the insets of Figure~\ref{fig:architecture}\textbf{b} for all the optimization algorithms. Despite achieving similar accuracy across the different algorithms, the training history reveals a faster convergence rate with heuristic methods. 

Although coupled waveguide arrays typically offer a linear wavelength response, the serial combination of them induces a richer wavelength dispersion. To illustrate this, Figure~\ref{fig:architecture}\textbf{c} shows the simulated transmission matrix of the waveguide array used in the layout design. The latter is presented in dB scale and illustrates the lack of features in the dispersion, as is the case in typical ring resonators. In turn, the experimental wavelength response of the device was characterized and shown in Figure~\ref{fig:architecture}\textbf{d}. In the latter, the device was programmed to target the 2143 permutation; that is, light from the sequence of input ports $\{1,2,3,4\}$ is entirely routed to the output sequence $\{2,1,4,3\}$. Indeed, these plots reveal a rich response that varies across different wavelengths. 

The latter results provide insight into the wavelength engineering of the device response. Of particular interest is the wavelength demultiplexing and filtering operation in PICs, which enables the separation of optical signals carried over a single channel into discrete wavelength channels, each directed to a particular designated output port for parallel processing. The present device can indeed be operated to perform filtering operations. To this end, the continuous-wave laser used in the experimental setup is swept over a wavelength range of 1500 nm to 1600 nm, which excites a single input port. This effectively generates a transfer matrix for each wavelength, sampling $N_{\lambda}$ wavelength points. The device is then trained to filter out up to five different wavelengths at the output ports one and two. This renders $N_{2}=1$ (one column) and $N_{1}=2$ (two rows) in the target matrices used in the figure of merit in Eq.~\eqref{L_multi}.

\begin{figure}[t]
   \centering
   \includegraphics[width=0.8\textwidth]{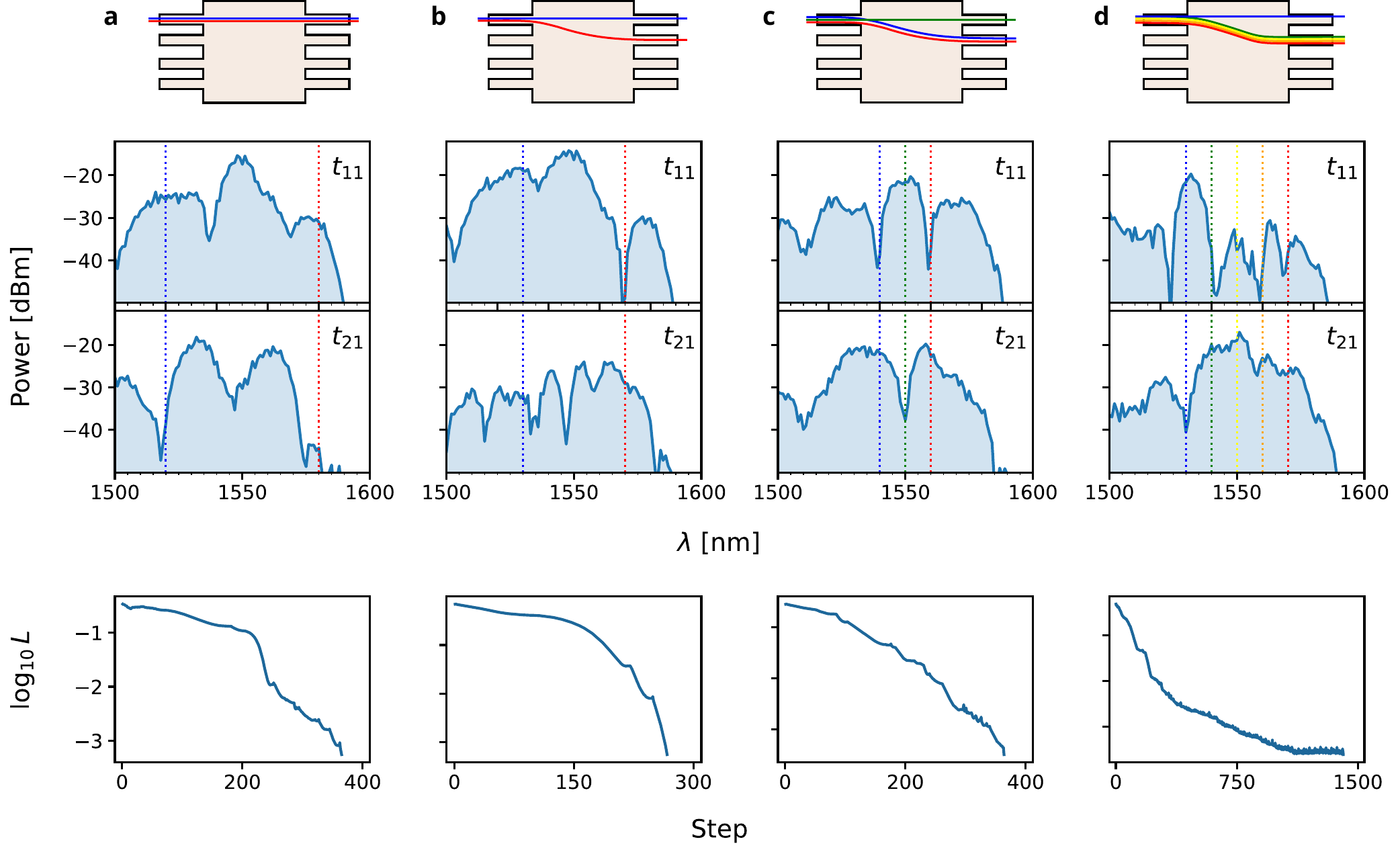}
   \caption{\textbf{Port-selective wavelength demultiplexing.} \textbf{a} Embedded 1x2 passthrough optimized for 1520 nm and 1580 nm through output 1, and not through output 2. \textbf{b} Embedded 1x2 demultiplexer optimized for 1530 nm through output 1 and 1570 nm through output 2. \textbf{c} Embedded bandpass filter optimized for 1550 nm through output 1, and 1540 and 1560 nm through output 2. \textbf{d} Embedded highpass filter optimized for 1530 nm through output 1, and 1540, 1550, 1560, and 1570 nm through output 2.
   }
   \label{fig:frequency_programmability}
\end{figure}

The following four demultiplexing configurations are used for training four different configurations as follows: (a) Optical signals at 1520 nm and 1580 nm are directed entirely to port one, with no power sent to port two (Figure~\ref{fig:frequency_programmability}\textbf{a}). (b) Optical signals at 1530 nm are routed to port one, while those at 1570 nm are directed to port two (Figure~\ref{fig:frequency_programmability}\textbf{b}). (c) Optical signals at 1540 nm and 1560 nm are routed to port two, while 1550 nm is steered to port one (Figure~\ref{fig:frequency_programmability}\textbf{c}). (d) Optical signals at 1540 nm, 1550 nm, 1560 nm, and 1570 nm are routed to port two, while 1530 nm is steered to port one (Figure~\ref{fig:frequency_programmability}\textbf{d}). The full wavelength sweep from 1500 nm to 1600 nm, along with the corresponding training history of the figure of merit, is shown in all cases. In the worst-performing scenario, the extinction ratio is at least 10 dB, which highlights the effectiveness of the device for demultiplexing tasks. 

\begin{figure}[t]
    \centering
    \includegraphics[width=0.8\textwidth]{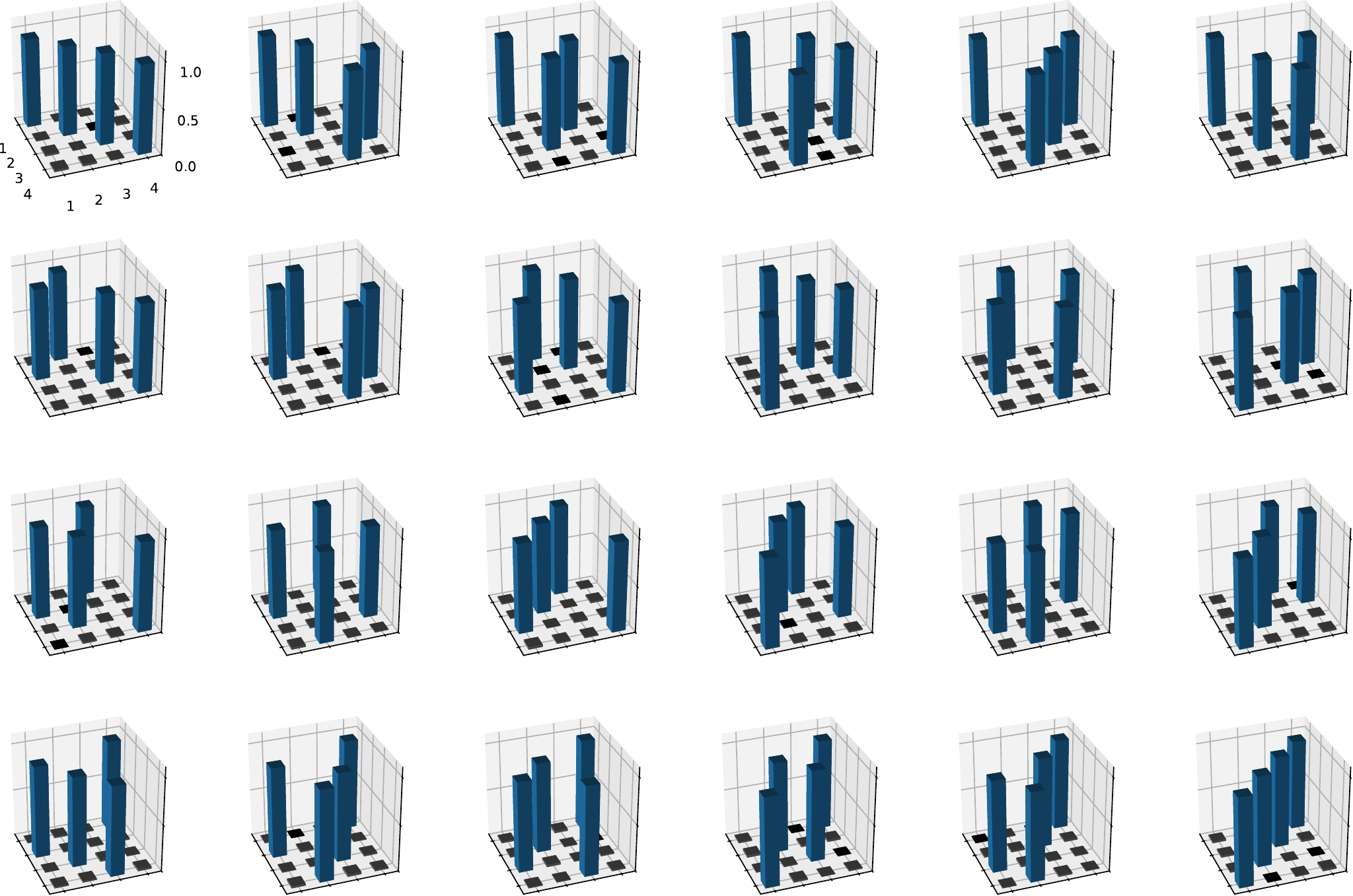}
    \caption{ \textbf{Single wavelength optimization of the device for representing sparse matrices.} Experimental measurements of the intensity transmission matrix of the device when programmed in-situ to represent $4\times 4$ permutation matrices at a single wavelength of $\lambda=1550$ nm. In this experiment, we considered the whole set of 24 permutation matrices.}
    \label{fig:Fig2}
\end{figure}

\section{Conclusion}
Our results demonstrate good performance of the PIC architecture as a programmable platform for optical switching and demultiplexing. The PIC is programmed on-the-fly through in-situ optimization by optimizing the current fed to the microheaters. By employing optimization algorithms such as gradient descent, hill climbing, and simulated annealing, we effectively calibrate the device to target the desired intensity matrix. During the process, high accuracy and fast convergence are achieved for sparse matrices. This programmable photonic chip exhibits versatility in achieving frequency demultiplexing and implementing sparse unitary matrices for optical switching. Several enhancements can be made to further improve the device capabilities. 

For completeness, a series of single-wavelength trainings is performed to assess the universality of unitary matrices at 1550 nm. Here, the four-dimensional representation of the permutation group is used to generate the target matrices. This renders $N!=24$ sparse unitary matrices comprising a unique one element per row and zero elsewhere~\cite{weyl1950theory}. The gradient-descent algorithm is chosen during the training of these sparse matrices, and the trained transmission matrices are shown in Figure~\ref{fig:Fig2}. The convergence of the device transmission matrix to the permutation matrix is achieved across all the elements of the permutation group with high accuracy, despite the device having three broken metal heaters. As discussed in previous works~\cite{Markowitz23Auto}, the device is effectively overparameterized and thus its performance is not compromised if some active elements are missing.

Our device is presented as a proof of concept to demonstrate the multiple operational tasks compatible with the device. However, device compactness can be further improved by utilizing deep trenches or employing phase shifter technologies that do not generate excess heat. This will reduce the space between phase elements in the layout design, which is by far the largest footprint in the device. Precise temperature control can also be employed to improve the speed of in-situ training. Furthermore, the design is simple and compatible with any other material platform and open-access foundries.




\section*{Funding}
This project is supported by the U.S. Air Force Office of Scientific Research (AFOSR) Young Investigator Program (YIP) Award\# FA9550-22-1-0189, by the Defense University Research Instrumentation Program (DURIP) Award\# FA9550-23-1-0539, and by the National Science Foundation Award\# CNS-2329021.

\section*{Data availability}
The datasets generated and analyzed during the current study are available from the corresponding 
author on reasonable request.

\bibliography{biblio}

\end{document}